\documentstyle[preprint,aps]{revtex}
\begin{document}
\draft
\title{Distribution of local density of states in disordered metallic
samples: logarithmically normal asymptotics.}
\author{ Alexander D. Mirlin}
\address{Institut f\"{u}r Theorie der Kondensierten Materie,
  Universit\"{a}t Karlsruhe, 76128 Karlsruhe, Germany}
\address{
and  Petersburg Nuclear Physics Institute, 188350 Gatchina, St.Petersburg,
Russia.}
\date{\today}
\maketitle
%\narrowtext
\tighten
\begin{abstract}
Asymptotical behavior of the distribution function of local density of
states (LDOS) in disordered metallic samples is studied with making
use of the supersymmetric
$\sigma$--model approach, in combination with the saddle--point
method. The LDOS distribution is found to have the logarithmically
normal asymptotics for quasi--1D and 2D sample geometry.
In the case of a quasi--1D sample,
the result is confirmed by the exact solution. In 2D case a perfect
agreement with an earlier renormalization group calculation is found.
In 3D the found asymptotics is of somewhat different type:
$P(\rho)\sim \exp(-\mbox{const}\,|\ln^3\rho|)$.

\end{abstract}
\narrowtext

\section{Introduction}
Mesoscopic fluctuations of various physical quantities in disordered
systems have been intensively investigated during the last years
\cite{ALW}. It
was understood that a whole distribution function is to be studied to
get the complete physical information concerning properties of a
disordered system. In particular, statistical fluctuations of local
quantities, such as eigenfunction intensity and local density of states
(LDOS) have attracted a considerable research interest recently.
Distrbutions of eigenfunction amplitudes and of LDOS are relevant for
description of fluctuations of tunneling conductance across a quantum
dot \cite{Alh}, of some atomic spectra properties \cite{gay}, and of a
shape of NMR line \cite{NMR}. The microwave cavity technique
\cite{eksp} allows to simulate a disordered electronic system and to
observe experimentally spatial fluctuations of the wave intensity
\cite{kud}.

Distribution of eigenfunction amplitudes
$|\psi_i^2(\bbox{r})|$, characterizes properly the spatial
fluctuations in a {\it closed} system with well defined energy
levels. On the other hand, in an {\it open} sample, connected to the
leads, the states are broadened in an energy space. In this case, it
is more appropriate to consider the distribution of LDOS,
$$
\rho(E,\bbox{r})=-\pi^{-1}\mbox{Im}G_R(\bbox{r},\bbox{r};E)\ ,
$$
where $G_R$ is the retarded Green function. Distribution of LDOS in
open metallic samples was studied by Altshuler, Kravtsov and Lerner
\cite{akl,lerner} for the case of two--
and $(2+\epsilon)$--dimensional systems. The consideration was based
on the derivation of the effective $\sigma$--model and its subsequent
renormalization group analysis, following the earlier ideas by Wegner
\cite{Weg}.  It was found that the LDOS
distribution is close to the Gaussian one but has slowly decaying
logarithmically normal (LN) asymptotics. When approaching the Anderson
transition in $(2+\epsilon)$ dimensions, the distribution was found to
cross over to the completely LN one. Altshuler and Prigodin
\cite{alpr} studied the LDOS distribution in strictly 1D chains and
also found the LN form of the distribution. It was conjectured on the
basis of this similarity \cite{akl,alpr} that even in a metallic
sample there is a finite probability to find ``almost localized''
eigenstates.

More recently, it was recognized \cite{mf,epr,epri,mf2} that the
statistical properties of various quantities characterizing
fluctuations of the wave functions can be very efficiently studied
with use of the supersymmetric approach. In particular, it was shown in
\cite{mf,mf2} that the distribution $P_y(y)$ of eigenfunction
amplitudes $y=V|\psi(\bbox{r_0}|^2$ (here $V$ is the system volume
introduced for normalization purposes) can be written in terms of the
supersymmetric $\sigma$--model in the following way. Let us define
the function $Y(Q_0)$ as
\begin{equation}
Y(Q_0)=\int_{Q(\bbox{r_0})=Q_0} DQ(r)
\exp\{-S[Q]\}\ ,
\label{1}
\end{equation}
where $S[Q]$ is the $\sigma$--model action
\begin{equation}
S[Q]=-\int d^dr\,\mbox{Str}\left[{\pi\nu D\over 4}(\nabla
Q)^2-\pi\nu\eta\Lambda Q\right]\ ,
\label{2}
\end{equation}
 $Q(\bbox{r})$ is a $4\times 4$ supermatrix field,
$\Lambda=diag\{1,1,-1,-1\}$,
$Str$ stands for the supertrace, $D$ is the diffusion constant,
$\eta$ is the level broadening
(imaginary frequency) and $\nu$ is the mean DOS \cite{note1}.
We do not go into details of the supersymmetric formalism here, which
can be found e.g. in \cite{efe,vwz,FMRev}.
For the invariance reasons, the function $Y(Q_0)$ turns out to be
dependent on the two scalar variables
$1\le \lambda_{1}<\infty $ and $ -1\le \lambda_{2} \le 1$, which are
eigenvalues of the ``retarded-retarded'' block of the matrix $Q_0$.
Moreover, in the limit $\eta \to 0$ (at a fixed finite value of the
system volume) only the dependence on $\lambda_1$ persists:
\begin{equation}
Y(Q_0)\equiv Y(\lambda_1,\lambda_2)\to Y_a(2\pi\nu\eta\lambda_1)
\label{3}
\end{equation}
The distribution of the eigenfunctions intensities is given by
\cite{mf,mf2}
\begin{eqnarray}
P_y(y)&=&{d^2 \over dy^2} Y_a(y/V) \nonumber\\
      &=& {d^2 \over dy^2} Y\left.(\lambda_1={y\over
2\pi\nu V\eta})\right|_{\eta\to 0}
\label{4}
\end{eqnarray}

The distribution function $P_\rho(\rho)$ of LDOS can be also expressed
through the function $Y(\lambda_1,\lambda_2)$ \cite{mf2,epr2}. For
convenience, we normalize $\rho$ by its mean value:
$\tilde{\rho}=\rho/\langle\rho\rangle$, where
$\langle \rho \rangle=\nu$, and write again $\rho$ instead
of $\tilde{\rho}$ henceforth. Then,
\begin{equation}
P(\rho)
=\frac{1}{4\pi}\frac{\partial^{2}}{\partial \rho^{2}}\left\{
\int_{\frac{\rho^{2}+1}{2\rho}}^{\infty}
d\lambda_{1}\overline{Y}(\lambda_1)
\left(\frac{2\rho}{\lambda_{1}-
\frac{\rho^{2}+1}{2\rho}}\right)^{1/2}\right\}\ ,
\label{5}
\end{equation}
where
\begin{equation}
\overline{Y}(\lambda_1)=\int_{-1}^{1}d\lambda_2
{Y(\lambda_1,\lambda_2)\over \lambda_1-\lambda_2}
\label{33}
\end{equation}
Let us note the important symmetry relation found in \cite{mf2}:
\begin{equation}
P(\rho^{-1})=\rho^3 P(\rho)
\label{5a}
\end{equation}
It follows from eq.(\ref{5}) and is completely independent on a
particular form of the function $Y(\lambda_1,\lambda_2)$. Obviously,
eq.(\ref{5a}) relates the small--$\rho$ asymptotical behavior of the
distribution $P_\rho(\rho)$ to its large-$\rho$ asymptotics.

Relations (\ref{4}) and (\ref{5}) gave the possibility to study the
distributions of eigenfunctions amplitudes and of LDOS in various
regimes. Exact calculation of the eigenfunction amplitudes
distribution in a quasi--1D sample of arbitrary length was performed
in \cite{mf}. LDOS distribution in a sample with given level width was
calculated in \cite{epr,epr2} for the 0D and in \cite{mf3} for the
quasi--1D case. In Ref.\cite{mf2}, the distributions were  studied
in the vicinity of the Anderson metal--insulator transition, where
$Y(Q)$ acquires the meaning of an order parameter function. In
Ref.\cite{fm1,fm2} the distribution $P_y(y)$ in a metallic sample was
studied by a perturbative method which is valid for not too large
$y$.

In a recent paper, Muzykantskii and Khmelnitskii \cite{mk} studied the
long--time asymptotics of the current relaxation in a disordered
metallic sample via the supersymmetric $\sigma$--model approach
combined with the saddle--point method. Fal'ko and Efetov \cite{falef}
employed the saddle--point approach suggested in \cite{mk} to study
the asymptotical behavior of the distribution of eigenfunctions
amplitude, $P_y(y)$, given by eq.(\ref{4}). In the present paper, we
apply this method to investigate the asymptotical behavior of the
distribution function of LDOS in an open sample (i.e. with leads being
attached). Precisely this distribution function was studied by
Altshuler, Kravtsov and Lerner \cite{akl,lerner}  in the renormalization
group approach, so that the direct comparison is possible. We find the
LN asymptotics of $P_\rho(\rho)$ in 2 dimensions, in  agreement with
the result
of \cite{akl}. As we discuss in Section 4, it is {\it not} completely
obvious that this agreement should have been expected.

In the case of a quasi--1D geometry, we also calculate the LDOS
distribution function using the exact solution of the 1D
$\sigma$--model \cite{rejaei}. In the metallic regime, the LN
asymptotics obtained by the saddle--point method is reproduced. In the
deeply localized regime, the distribution function $P_\rho(\rho)$
takes a completely LN form.

\section{Asymptotical behavior of the LDOS distribution function from
the saddle--point approximation.}

We consider a disordered sample in the weak localization regime, which
means that
\begin{equation}
\xi=\int_{L^{-1}}^{l^{-1}} {d^d q\over (2\pi)^d} {1\over \pi\nu
Dq^2}\ll 1
\label{b1}
\end{equation}
where $\xi$ is the usual parameter of the perturbation theory
\cite{akl}. In this situation, the distribution of normalized local
DOS $\rho$ is mostly cocncentrated in a narrow Gaussian peak around
its mean value $\rho=1$, with the width
$\langle(\rho-1)^2\rangle\sim\xi\ll 1$ \cite{akl,lerner}. This
close-to-Gaussian shape of $P_\rho(\rho)$ holds however in the region
$|\rho-1|\ll 1$ only. We will consider in contrast the ``tails'' of
the distribution, $\rho\ll 1$ and $\rho\gg 1$, where a much slower
decay of $P_\rho(\rho)$ will be found.

As is seen from eq.(\ref{5}), the LDOS distribution
$P_\rho(\rho)$ is determined by the function $Y(\lambda_1,\lambda_2)$
at $\lambda_1\ge{\rho^2+1 \over \rho}$. Therefore, in the asymptotical
region $\rho\gg 1$, only the form of $Y(\lambda_1,\lambda_2)$ at large
values of $\lambda_1\ge\rho/2\gg 1$ is relevant. In this region, and
under the condition (\ref{b1}),
$Y(\lambda_1,\lambda_2)$ can be found with exponential accuracy with
 use of the saddle--point method \cite{mk,falef}. We consider the case of
an open samle and neglect any inelastic processes in the bulk of the
sample. The energy levels are then broadened solely due to the
possibility for a particle to escape to the leads, and the action is
given by eq.(\ref{2}) at $\eta=0$,
\begin{equation}
S[Q]=-\int d^dr\,\mbox{Str}{\pi\nu D\over 4}(\nabla
Q)^2
\label{6}
\end{equation}
The saddle point equations have the same form as in Ref.\cite{mk} for
$\omega=0$:
\begin{equation}
\triangle\theta_1=\triangle\theta_2=0,
\label{7}
\end{equation}
where $\cosh\theta_1=\lambda_1$; $\cos\theta_2=\lambda_2$, and
$\triangle$ is the Laplace operator. Assuming the perfect coupling to
the leads, the corresponding boundary condidtion takes the form
$Q|_{\mbox{leads}}=\Lambda$, or in other words,
\begin{equation}
\theta_1,\ \theta_2|_{\mbox{leads}}=0
\label{8}
\end{equation}
Another boundary condition which follows from the definition (\ref{1})
of the function $Y(\lambda_1^{(0)},\lambda_2^{(0)})$, fixes the value
of $\theta_1$, $\theta_2$ in the observation point $\bbox{r_0}$,
where the LDOS distribution is studied:
\begin{equation}
\cosh \theta_1(\bbox{r_0})=\lambda_1^{(0)}\equiv \cosh\theta_1^{(0)}
\ ;\qquad
\cos \theta_2(\bbox{r_0})=\lambda_2^{(0)}\equiv \cos\theta_2^{(0)}
\label{9}
\end{equation}

We start from the case of a quasi--1D sample. Let the observation
point be at $x=0$, whereas the sample edges at $x=-L_-$ and $x=L_+$,
so that the sample length is $L=L_-+L_+$. The solution of the
saddle--point equation (\ref{7}) with the boundary conditions
(\ref{8}), (\ref{9}) is
\begin{equation}
\theta_1(x)=\left\{\begin{array}{l}
\theta_1^{(0)}\left(1-{x\over L_+}\right)\ ;\qquad x>0 \\
\theta_1^{(0)}\left(1-{|x|\over L_-}\right)\ ;\qquad x<0\ ;
\end{array}\right.
\label{10}
\end{equation}
and similarly for $\theta_2$. The action on the saddle point solution
is given by
\begin{equation}
S={\pi\nu D\over 2}\int[(\nabla\theta_1)^2 +(\nabla\theta_2)^2]\,d^dr=
{\pi\nu D\over 2}[\theta_1^{(0)2}+\theta_2^{(0)2}]\left({1\over L_+}
+ {1\over L_-}\right)
\label{11}
\end{equation}
We find therefore that in the region $\lambda_1^{(0)}\gg1$,
\begin{equation}
Y(\lambda_1^{(0)},\lambda_2^{(0)})\sim e^{-S}\sim
\exp\left\{-{\pi\nu D\over
2}\left({1\over L_+}+ {1\over L_-}\right)
\ln^2(2\lambda_1^{(0)})\right\},
\label{11a}
\end{equation}
with exponential accuracy. Dependence of $Y$ on $\lambda_2^{(0)}$ is
not important within this accuracy, since it gives simply a prefactor
after the $\lambda_2$--integration in eq.(\ref{33}).
Therefore, according to eq.(\ref{5}) the
asymptotics of the LDOS distribution function at $\rho\gg 1$  is
\begin{equation}
P_\rho(\rho)\approx \overline{Y}(\lambda_1^{(0)}=\rho/2)
\sim\exp\left\{-{\pi\nu D\over
2}\left({1\over L_+}+ {1\over L_-}\right)\ln^2 \rho\right\}\ ,
\label{12}
\end{equation}
up to a preexponential factor. According to the symmetry relation
(\ref{5a}), eq.(\ref{12}) is equally describes the small--$\rho$
asymptotics of $P_\rho(\rho)$ at $\rho\ll 1$.

The found LN asymptotics of the LDOS distribution  in an {\it
open} sample should be contrasted to the ``square-root-exponential''
behavior of $P_\rho(\rho)$ and $P_y(y)$ in a {\it closed} sample
\cite{mf,FMRev,mf3,fm2}. Let us show  how the difference appears
in the present saddle--point approach. In a closed sample, the
distribution of LDOS, $P_\rho(\rho)$ is defined only if a finite level
width $\eta$ in the action, eq.(\ref{2}), is introduced. It may
account for some inelastic scattering processes in the bulk of the
sample. The saddle point equation acquires then the same form, as in
ref.\cite{mk}:
\begin{equation}
D\triangle\theta_1-2\eta\sinh\theta_1=0
\label{13}
\end{equation}
Eq.(\ref{13}) should be supplemented by the boundary conditions
\begin{equation}
\nabla_{\bbox{n}}\theta_1|_{\mbox{leads}}=0\ ,
\label{8a}
\end{equation}
where $\nabla_{\bbox{n}}$ is the normal derivative, and by
the condition (\ref{9}) in the observation point.

Let us first assume that  the saddle-point solution satisfies the
condition $\theta_1(x)\gg 1$ throughout the sample. It will be seen
below that this assumption is valid if $E_c/\eta\gg 1$, where
$E_c=D/L^2$ is the Thouless energy. Then $\sinh\theta_1$ in
eq.(\ref{13}) can be approximated by $e^{\theta_1}/2$, and the
solution for $x>0$ is found to be
\begin{equation}
e^{-\theta_1(x)}={\eta\over DC}\left\{1+\cos\left[\sqrt{C}(L_+-x)
\right]\right\}\ ,
\label{13a}
\end{equation}
where the constant $C$ is defined from the condition
\begin{equation}
L_+\sqrt{C}=\pi-\arccos\left(1-{CD\over 2\eta\lambda_1^{(0)}}\right)
\label{13b}
\end{equation}
If $\lambda_1^{(0)}\gg E_c/\eta$, the solution (\ref{13a}),
(\ref{13b}) takes the form
\begin{equation}
e^{-\theta_1(x)}={\eta\over D}\left(L_+\over\pi\right)^2\left\{
1-\cos\left[ {\pi\over L_+}\left(x+\sqrt{ {D\over\eta\lambda_1^{(0)} }}
\right)\right]\right\}
\label{13c}
\end{equation}
The action
\begin{equation}
S\simeq\int dx\left\{ {\pi\nu D\over 2}(\theta_1')^2+\pi\nu\eta
e^{\theta_1} \right\}
\label{13d}
\end{equation}
is then found to be dominated by the vicinity of the observation
point, where eq.(\ref{13c}) can be reduced to
\begin{equation}
e^{-\theta_1(x)}={1\over 2\lambda_1^{(0)}}\left( 1+ x \sqrt{
{\eta\lambda_1^{(0)}\over D} }\right )^2
\label{13e}
\end{equation}
Substituting (\ref{13e}) in (\ref{13d})  and
taking into account also an analogous contribution of the $x<0$
region, we get
\begin{equation}
S\simeq 8\pi\nu\sqrt{ D\eta\lambda_1^{(0)}}
\label{13f}
\end{equation}
In the opposite case $1\ll \lambda_1^{(0)}\ll E_c/\eta$, the solution
(\ref{13a}), (\ref{13b}) is reduced to
\begin{equation}
\theta_1(x)=\theta_1^{(0)}-{\eta\lambda_1^{(0)}\over D} (2x L_+-x^2)
\label{13g}
\end{equation}
The corresponding action is
\begin{equation}
S\simeq 2\pi\nu\eta\lambda_1^{(0)}L
\label{13h}
\end{equation}
Combining (\ref{13f}) and (\ref{13h}), we find therefore for
$E_c/\eta\gg 1$
\begin{equation}
P_\rho(\rho)\sim Y(\lambda_1^{(0)}=\rho/2)\sim\left\{
\begin{array}{ll}
\exp\{-4\pi\nu\sqrt{2D\eta\rho} \}\ , &\quad \rho\gg{E_c\over\eta}\gg
1\\
\exp\{-\pi\nu\eta L\rho\}\ , &\quad {E_c\over\eta}\gg\rho\gg 1
\end{array}\right.
\label{13i}
\end{equation}
If $E_c/\eta\lesssim 1$, one has to use the exact equation (\ref{13}),
which has the following solution:
\begin{equation}
x(\theta_1)=\left( {D\over 4\eta}\right)^{1/2}
\int_{\theta_1}^{\theta_1^{(0)}} {d\vartheta\over
[\cosh\vartheta-\cosh\theta_1(L)]^{1/2} }\ ,
\label{13k}
\end{equation}
where $\theta_1(L)$ is defined from the condidtion
\begin{equation}
L=\left( {D\over 4\eta}\right)^{1/2}
\int_{\theta_1(L)}^{\theta_1^{(0)}} {d\vartheta\over
[\cosh\vartheta-\cosh\theta_1(L)]^{1/2} }\ ,
\label{13l}
\end{equation}
We find that for $E_c\lesssim\eta$ the solution (\ref{13k}) in the
vicinity of the observation point $x\ll\sqrt{D/\eta}$ has exactly the
form (\ref{13e}) for all $\lambda_1^{(0)}\gg 1$, so that the action is
given by eq.(\ref{13f}). Thus, in this case the first of the
asymptotics (\ref{13i}) is applicable
\begin{equation}
P_\rho(\rho)\sim \exp\{-4\pi\nu\sqrt{2D\eta\rho} \}\ ,\quad
{E_c\over\eta}\lesssim 1\ ,\quad \rho\gg 1
\label{13m}
\end{equation}
Note that eq.(\ref{13m}) has exactly the same form as the asymptotics
of $P_\rho(\rho)$ in the infifnitely long wire
\cite{mf3}. Eq.(\ref{13i}) is also in agreement with the asymptotical
behavior of the distribution function $P_y(y)$ of eigenfunctions
intensities \cite{fm1}:
\begin{equation}
P_y(y)\sim \exp\{-4\sqrt{2\pi\nu D y/L}\}\ ;\quad y\gg {2\pi\nu D\over
L}
\label{13n}
\end{equation}
which is immediately reproduced from eqs.(\ref{4}), (\ref{13n}).

We return now to the case of an open sample and consider 2D
geometry. As was suggested in \cite{mk}, we suppose the sample to be a
disk of a radius $L$ surrounded by a perfectly conducting media, with
the observation point located in the center of the disk. We will find
the result to be logarithmically dependent on $L$, that justifies its
application to a sample of more general 2D geometry with a
characteristic size $L$.

Eq.(\ref{7}) for $\theta_1(\bbox{r})$ with the boundary conditions
(\ref{8}), (\ref{9}) take now the form
\begin{eqnarray}
&&\theta_1''(r)+{\theta_1'(r)\over r}=0 \label{20}\\
&&\theta_1(l_*)=\theta_1^{(0)}                         \label{21}\\
&&\theta_1(L)=0                 \label{22}
\end{eqnarray}
The boundary condition (\ref{21}) has to be put
not in the observation point $r=0$, but rather at certain distance
from it $r=l_*$ \cite{mk}. This is because the
$\sigma$--model approximation breaks down for momenta $q> l^{-1}$,
where $l$ is of order of the mean free path.
We will specify the value of $l_*$ below.
The solution of eq.(\ref{20}) with the boundary conditions
(\ref{21}), (\ref{22}) reads
\begin{equation}
\theta_1(r)=\theta_1^{(0)}-{\theta_1^{(0)}\over\ln L/l_*}\ln r/l_*\ ,
\qquad l_*<r<L\ ,
\label{23}
\end{equation}
or, for $\lambda_1(r)=\cosh\theta_1(r)\simeq{1\over
2}e^{\theta_1(r)}$,
\begin{equation}
\lambda_1(r)=\lambda_1^{(0)}\left({l_*\over
r}\right)^{\theta_1^{(0)}\ln^{-1} (L/l_*)}
\label{24}
\end{equation}
The condition of applicability of the $\sigma$--model (diffusive)
approximation is \cite{mk} $d\theta_1/dr<1/l$. It  leads to the
following equation for the cut-off scale $l_*$
\begin{equation}
{1\over l} =
\left. {d\theta_1\over dr}\right|_{r=l_*}
\equiv {\theta_1^{(0)}\over\ln (L/l_*)} {1\over l_*}
\ ,
\label{34z}
\end{equation}
The action (\ref{6}) is now given by
\begin{eqnarray}
S&\simeq&{\pi\nu D\over 2}\int d^2r(\nabla\theta_1)^2=\pi^2\nu
D\int_{l_*}^L  dr\, r[\theta_1'(r)]^2 \nonumber \\
&=& \frac{\pi^2\nu D \theta_1^{(0)2}}{\ln L/l_*}\ .
\label{25}
\end{eqnarray}
Therefore, the LDOS distribution has  the  asymptotics
\begin{equation}
P_\rho(\rho)\sim\exp\left\{-\frac{\pi^2\nu D \ln^2\rho}
{\ln (L/l_*)}\right\}\ ,
\label{26}
\end{equation}
where $l_*$ satisfies the condition
\begin{equation}
l_*=l {\ln\rho\over \ln (L/l_*) }
\label{25a}
\end{equation}
The present consideration is meaningful provided $l_*\ll L$, which holds
if $\ln\rho\ll L/l$. Under this condidtion, eq.(\ref{25a}) yields
\begin{equation}
l_*\simeq l{\ln\rho\over\ln\left( {L\over l\ln\rho}\right)}
\label{25b}
\end{equation}
Taking into account the logarithmic dependence of the exponent in
(\ref{26}) on $l_*$ and neglecting the corrections of the
$\ln(\ln\ldots)$ type, we can approximate $l_*$ in eq.(\ref{26}) by
$l$. The result is of the LN form,
in  exact agreement with the asymptotical behavior for
$P_\rho(\rho)$ found by the renormalization group method in
\cite{akl}. We discuss this agreement in more detail in Conclusion.
Let us also note that in contrast to the 1D case, in 2D the
asymptotical form of the distribution $P_\rho(\rho)$ in an open
sample, eq.(\ref{26}), is similar to that of $P_y(y)$ in a closed
sample \cite{falef}.

Proceeding in the same way in 3D case, we get the equation
\begin{equation}
\theta_1''(r)+{\theta_1'(r)\over r^2}=0\ , \label{27}
\end{equation}
with the same boundary conditions (\ref{21}), (\ref{22}). The solution
has the form
\begin{equation}
\theta_1(r)=\theta_1^{(0)}{l_*\over r}\ ,\qquad l_*<r<L\
\label{28}
\end{equation}
The condition (\ref{34z}) for $l_*$ now yields $l_*\simeq
l\theta_1^{(0)}$, and the action is  estimated as
\begin{eqnarray}
S&=&2\pi^2\nu D\int_{l_*} dr\, r^2[\theta_1'(r)]^2 \nonumber \\
&=& 2\pi^2\nu D l_*\theta_1^{(0)2} \nonumber\\
&=& 2\pi^2\nu D l \theta_1^{(0)3} \nonumber\\
&=&\pi g(l)\theta_1^{(0)3}\ , \label{29}
\end{eqnarray}
where $g(l)=2\pi^2\nu Dl$ is the conductance of a cube of the size
$l$. Therefore, the LDOS distribution asymptotics is
\begin{equation}
P_\rho(\rho)\sim\exp\{-g(l)\ln^3\rho\}\ ,
\label{30}
\end{equation}
 Let us note that $l$ is of order of mean
free path, so that it is defined up to a numerical coefficient of
order unity. This ambiguity does not affect the leading order behavior
of the exponent in eq.(\ref{26}) for the 2D case, due to its
logarythmic dependence on $l$. On the other hand, in 3D case, the
exponent in (\ref{30}) is proportional to $l$. To fix the
corresponding numerical coefficient, one should go beyond the
long-wave-length $\sigma$--model approximation and consider the
problem in the region $r\lesssim l_*$.

To conclude this section, we have shown that the LDOS distribution
$P_\rho(\rho)$ has logarithmically normal asymptotics (\ref{12}),
(\ref{26}) at $\rho\gg 1$ and $\rho\ll 1$
for quasi--1D and 2D samples, whereas for 3D systems a somewhat
different result (\ref{30}) was found.
In the next section, we confirm this result for the quasi-1D case by
the exact solution of the problem.

\section{LDOS distribution in a quasi--1D sample: exact solution.}

In the case of a quasi--1D geometry of a sample, evaluation of the
functional integral in (\ref{1}) can be reduced to the solution of an
evolution equation of the diffusion type for the function
$Y_1(\lambda_1,\lambda_2;x)$ \cite{efe}:
\begin{eqnarray}
&& 2\pi\nu D{\partial Y_1 \over\partial x}=\left[{1\over J}
{\partial\over\partial\theta_1}J{\partial\over\partial\theta_1}
+{1\over J}
{\partial\over\partial\theta_2}J{\partial\over\partial\theta_2}
\right] Y_1\ ;   \nonumber\\
&& J(\theta_1,\theta_2)=\frac{|\sinh\theta_1\sin\theta_2|}
{(\cosh\theta_1-\cos\theta_2)^2}
\label{30a}
\end{eqnarray}
For the general case of a finite quasi--1D sample with distances
$L_-$, $L_+$ from the observation point to the edges, the function
$Y(\lambda_1,\lambda_2)$ defined in the preceding section is given by
the product
\begin{equation}
Y(\lambda_1,\lambda_2)=Y_1(\lambda_1,\lambda_2;L_-)
Y_1(\lambda_1,\lambda_2;L_+)
\label{31}
\end{equation}
For simplicity, we will concentrate on the case of a semi-infinite
sample $L_+=\infty$, $L_-=L$. In this case
$Y(\lambda_1,\lambda_2)=Y_1(\lambda_1,\lambda_2;L)$, and we will omit
the subscript ``1'' in the notation ``$Y_1$''. Results for the general
case will be presented in the end of the section. Solution of
eq.(\ref{30a}) with the boundary condition corresponding to a perfectly
conducting lead was recently found to be \cite{rejaei}:
\begin{eqnarray}
&&Y(\lambda_1,\lambda_2;L)=1-(\lambda_1-\lambda_2)
\sum_{m=0}^{\infty}\int_0^\infty dk\,
c_{mk} P_m(\lambda_2)P_{-1/2+ik/2}(\lambda_1)e^{-\epsilon_{mk}t}\ ;
\nonumber \\
&&t={L\over 2\pi\nu D}\ ;\qquad
c_{mk}={(2m+1)k\tanh(\pi k/2)\over (2m+1)^2+k^2}\ ;\qquad
\epsilon_{mk}=(m+1/2)^2+k^2/4
\label{32}
\end{eqnarray}
Substituting eq.(\ref{32}) in (\ref{33}), we find
\begin{equation}
\overline{Y}(\lambda_1)=\ln{\lambda_1+1\over\lambda_1-1}-
2\int_0^\infty{dk\over 1+k^2}k\,\tanh(\pi k/2) P_{-1/2+ik/2}(\lambda_1)
e^{-t(k^2+1)/4}
\label{34}
\end{equation}
The first term in eqs.(\ref{32}) and (\ref{34}) corresponds to the
limit of a closed system with no level broadening, in which case
$Y(\lambda_1,\lambda_2)=1$ and $P_\rho(\rho)=\delta(\rho)$. It can be
really proven by direct substitution of the first term of
eq.(\ref{34}) into eq.(\ref{5}) that the resulting contribution to
$P_\rho(\rho)$ is equal to zero at any $\rho\ne 0$. To evaluate the
second term, we use the asymptotic expression for the Legendre
functions $P_{-1/2+i\gamma}(z)$ at $z\gg 1$ \cite{ryzhik}:
\begin{equation}
P_{-1/2+i\gamma}(z)\simeq{1\over\sqrt{2\pi z}}\left\{e^{i\gamma\ln 2z}
{\Gamma(i\gamma)\over\Gamma(1/2+i\gamma)}+e^{-i\gamma\ln 2z}
{\Gamma(-i\gamma)\over\Gamma(1/2-i\gamma)}\right\}
\label{34a}
\end{equation}
This allows us to reduce the second term in (\ref{34}) to the form
\begin{equation}
\overline{Y}_(2)(\lambda_1)=-
4\int_{-\infty}^\infty{dk\over 1+k^2}{1\over\sqrt{2\pi\lambda_1}}
{\Gamma(1/2-ik/2)\over \Gamma(-ik/2)}e^{i{k\over 2}\ln 2\lambda_1}
e^{-t(k^2+1)/4}
\label{35}
\end{equation}
 At $\ln\lambda_1\gg 1$ the integral can be evaluated via the saddle
point method. The saddle point is
$k=i{\ln 2\lambda_1\over t}$ and yields the following contribution:
\begin{equation}
\overline{Y}_{\mbox{s.p.}}(\lambda_1)\simeq -4\sqrt{{2\over t\lambda_1}}
{1\over 1-\left({\ln 2\lambda_1\over t}\right)^2}
{\Gamma\left({1\over 2} + {\ln 2\lambda_1\over 2t}\right)\over
\Gamma\left({\ln 2\lambda_1\over 2t}\right)}
\exp\left\{-{t\over 4}-{\ln^2 2\lambda_1\over 4t}\right\}
\label{36}
\end{equation}
In fact, when deriving eq.(\ref{36}), we should shift the integration
contour in the complex plane in order that it would pass through the
saddle point. However, if $\ln 2\lambda_1>t$, the contour crosses then
the pole of the integrand, $k=i$. Evaluating this pole contribution,
we find that in cancels exactly the first term in eq.(\ref{32}),
(\ref{34}). Thus, we get
\begin{equation}
\overline{Y}(\lambda_1)\simeq \ln{\lambda_1+1\over\lambda_1-1}
\vartheta(\ln 2\lambda_1-t)+
\overline{Y}_{\mbox{s.p.}}(\lambda_1)
\label{37}
\end{equation}
where $\vartheta(x)$ is the step function and
$\overline{Y}_{\mbox{s.p.}}$ is given by eq.(\ref{36}). Now we
substitute
this result into the formula (\ref{5}) for the LDOS distribution. The
leading contribution at $\rho\gg 1$ is found to be given by the second
term, $\overline{Y}_{\mbox{s.p.}}(\lambda_1)$, in (\ref{37}). It can
be found by noticing that the factor
$$
{1\over 1-\left({\ln 2\lambda_1\over t}\right)^2}
{\Gamma\left({1\over 2} + {\ln 2\lambda_1\over 2t}\right)\over
\Gamma\left({\ln 2\lambda_1\over 2t}\right)}
$$
varies logarithmically slow with $\lambda_1$, and can be simply taken
at $\lambda_1=\rho/2$. The remaining integral is of the form
\begin{eqnarray}
&&\int_{\rho/2}d\lambda_1{1\over\sqrt{\lambda_1(\lambda_1-\rho/2)}}
\exp\left\{-{t\over 4}-{\ln^2 2\lambda_1\over 4t}\right\} \nonumber\\
&&=\int_1^\infty{dy\over\sqrt{y(y-1)}}\exp\left\{-{t\over 4}-
{\ln^2\rho\over 4t}- {\ln\rho\ln y\over 2t}-{\ln^2 y\over 4t}\right\}
\nonumber\\
&& \simeq B\left({1\over 2},{\ln\rho\over 2t}\right)
\exp\left\{-{t\over 4}-{\ln^2\rho\over 4t}\right\}
\label{39}
\end{eqnarray}
where B(x,y) is the Euler's beta function. When writing the last line,
we assumed that $\ln\rho>\sqrt{t}$. It is not a restriction in the
metallic (short sample) case, $t<1$, and includes the region of
characteristic LDOS, $\ln\rho\sim t$, in the insulating regime, $t\gg
1$.

Collecting now all the remaining factors in eqs.(\ref{5}), (\ref{36}),
we finally get the distribution of LDOS:
\begin{equation}
P_\rho(\rho)\simeq{1\over 2\rho\sqrt{\pi t}}\exp\left\{-{1\over
4t}(t+\ln\rho)^2\right\}
\label{40}
\end{equation}
Let us remind that eq.(\ref{40}) has been derived under the assumption
$\rho\gg 1$. An additional condition $\ln\rho\gg\sqrt{t}$ in the
insulating regime is not very restrictive, since the characteristic
scale of $\ln\rho$ in this case is $\ln\rho\sim t$, as is seen from
(\ref{40}). Moreover, calculating the integral (\ref{39}) in the
opposite case $\ln\rho\ll t$, we arrive at the same LN behavior as in
eq.(\ref{40}); only the prefactor is different:
\begin{equation}
P_\rho(\rho)\simeq{\pi|\ln\rho|\over 4\rho t}\exp\left\{-{1\over
4t}(t+\ln\rho)^2\right\}
\label{41}
\end{equation}
Therefore, we have found the LN asymptotic behavior (\ref{40}),
(\ref{41}) of the LDOS distribution $P_\rho(\rho)$ ar $\rho\gg 1$. The
symmetry relation (\ref{5a}) allows us to extend its validity to the
region of $\rho\ll 1$ as well. Note that eqs.(\ref{40}, (\ref{41})
completely preserve their form under the transformation (\ref{5a}).
Eqs.(\ref{40}), (\ref{41}) are in full agreement with the results of
the saddle point approximation of section 2.

Let us remind theat the above results have been derived for the case
of a semiinfinite sample: $t_+=\infty$, $t_-=t$, where
$t_\pm=L_\pm/2\pi\nu D$. Now we consider briefly the general case of
finite $t_+,\ t_-$. In this case the function $Y(\lambda_1,\lambda_2)$
determining the LDOS distribution is defined by eq.(\ref{31}) with
$Y_1(\lambda_1,\lambda_2;L)$ given by eq.(\ref{32}). Using the above
analysis of the asymptotic behavior of $Y_1$ at $\lambda_1\gg 1$, we find
with an additional condition $\ln 2\lambda_1>t_+,t_-$,
\begin{equation}
\overline{Y}(\lambda_1;t_+,t_-)={\lambda_1\over 2}
\overline{Y}_{\mbox{s.p.}}(\lambda_1,t_+)
\overline{Y}_{\mbox{s.p.}}(\lambda_1,t_-)\ ,
\label{42}
\end{equation}
with $\overline{Y}_{\mbox{s.p.}}(\lambda_1,t)$ given by
eq.(\ref{36}). This leads to the following result for the LDOS
distribution:
\begin{equation}
P_\rho(\rho)\simeq\frac
{ F\left({\ln\rho\over 2t_+}\right)F\left({\ln\rho\over 2t_-}\right) }
{F\left({\ln\rho\over 2t_+} + {\ln\rho\over 2t_-} -1/2\right) }
{1\over 4\rho\sqrt{\pi t_+ t_-}          }
\exp\left\{ -{1\over 4}(t_+ +t_-)-{1\over 4}
\left({1\over t_+}+{1\over t_-}\right) \ln^2\rho\right\}
\label{43}
\end{equation}
where
$$
F(x)={\Gamma(x-1/2)\over(x+1/2)\Gamma(x)}
$$
Eq.(\ref{43}) holds in the metallic regime $t_+,t_-<1$ for all
$\rho\gg 1$ and is again in agreement with the result of the previous
section, eq.(\ref{12}). In the insulating regime eq.(\ref{43}) holds
provided the additional condition $\ln\rho>t_-,t_+$ is satisfied and
represents therefore the far asymptotics. In this case, there is also an
intermediate regime $1<\ln\rho<t_+$,  where we assumed for
definitness $t_-<t_+$. Then $Y(\lambda_1,\lambda_2,t_+)$ can be
approximated by 1, and we find $P_\rho(\rho)$ given by eq.(\ref{40})
with $t=t_-$. In all the cases, the equality (\ref{5a}) allows to get
the asymptotical behavior in the region of small LDOS, $\rho\ll 1$.

In conclusion of this section, let us note that in the insulating
regime, $t_-,t_+\gg1$, the obtained results for $P_\rho(\rho)$
are completely analogous to what was found by Altshuler and Prigodin
\cite{alpr} for the case of a strictly 1D sample by Berezinskii
technique.  This confirms once more the general conjecture
\cite{FMRev} that the statistics of smooth envelopes of the wave
functions in 1D and quasi--1D samples are equivalent.

\section{Discussion and conclusion.}

In this article, we have studied the asymptotical behavior of the LDOS
distribution function $P_\rho(\rho)$ in disordered metallic samples
at $\rho\ll 1$ and $\rho\gg 1$.
For this purpose, we  used the supermatrix $\sigma$--model
approach, which allows to express the distribution $P_\rho(\rho)$ in
terms of the function $Y(\lambda_1,\lambda_2)$ defined as a certain
integral over the supermatrix field. In the asymptotical regime,
 $\rho\ll 1$ and $\rho\gg 1$, this integral can be
estimated via the saddle-point method, leading to the log-normal
asymptotics of $P_\rho(\rho)$ for  quasi--1D and  2D  sample
geometries, and to the result (\ref{30}) in the 3D case.

In quasi--1D the result can be also found from the exact solution, which
indeed yields the LN asymptotical behavior in the case of a short
(metallic) sample. In the opposite case of a sample much longer than
the localization length, the whole distribution function
$P_\rho(\rho)$ has the log-normal form. This is completely analogous
to the form of the LDOS distribution in a {\it strictly} 1D sample
in the strongly localized regime studied in \cite{alpr} by the
Berezinskii technique.

In 2D, the obtained asymptotics of the LDOS distribution is in full
agreement with the result of renormalization group treatment
\cite{akl}. This agreement is highly non-trivial, for the following
reason. The RG treatment is based on a resummation of the perturbation
theory expansion and can be equally well performed within the replica
(bosonic or fermionic) or supersymmetric formalism. At the same time,
the present approach based on the supersymmetric formalism relies
heavily on the topology of the saddle-point manifold combining
non-compact ($\lambda_1$) and compact ($\lambda_2$) degrees of
freedom. The asymptotic behavior of $P_\rho(\rho)$ at $\rho\gg 1$ and
$\rho\ll 1$ is determined by the region $\lambda_1\gg 1$ which is very
far from the ``perturbative'' region of the manifold $Q\simeq\Lambda$
(i.e. $\lambda_1,\lambda_2\simeq 1$). It is well known \cite{vz}
that for the problem
of energy level correlation, the replica approach is not able to
reproduce the correct result (which can be obtained by the
supersymmetry method) even if all orders of perturbation theory are
taken into account. This is related to the fact that within the
replica method the topology of the saddle-point manifold is not
properly reflected. This might suggest a conclusion that any result
based on a resummation of the perturbative expansion (which can be
therefore obtained within the replica approach) is incorrect for the
same reason. The agreement of the results of supersymmetric and
renormaliztion group treatments of the LDOS distribution shows however
that this conclusion would be wrong. Apparently, this problem is of a
diferent kind, as compared to the level correlation one, so that the
replica trick combined with the RG is able to reproduce the correct
result. The difference seems to be the following: in the
case of level correlator both compact and non-compact sectors in the
supersymetric formulation are equally important, whereas for the
present problem only the non-compact sector was essential, with
compact one playing an auxiliary role. Let us note that the same
situation appears in the vicinity of the Anderson transition
\cite{zirn,mf2}
where the function $Y(\lambda_1,\lambda_2)$
acquires a role of the order parameter function and dependes on the
non-compact variable $\lambda_1$ only. The above agreement obtained
for the LDOS distribution provides therefore  support to other
results \cite{akl}
obtained with making use of the renormalization group approach
and $2+\epsilon$ expansion.

This work was supported by SFB 195 der Deutschen Forschungsgemeinschaft.


\begin{references}
\bibitem{note1} We consider the technically simplest unitary symmetry
class throughout the paper. The results are expected however
to be completely analogous for the orthogonal and symplectic symmetry
classes.
\end{references}
\end{document}